\pdfoutput=1

\documentclass[sigconf]{acmart}

\usepackage{booktabs} 
\usepackage[german,american]{babel}
\usepackage[T1]{fontenc}
\usepackage[latin1]{inputenc} 
\usepackage{amsmath}
\usepackage{multirow}

\newcommand{\Ni}{(1)~}
\newcommand{\Nii}{(2)~}
\newcommand{\Niii}{(3)~}
\newcommand{\Niv}{(4)~}
\newcommand{\Nv}{(5)~}
\newcommand{\Nvi}{(6)~}

\newcommand{\query}[1]{{\texttt{\small #1}}}
\newcommand{\interpretation}[1]{$\langle$\query{#1}$\rangle$}

\begin{document}
\fancyhead{}

\copyrightyear{2022}
\acmYear{2022}
\setcopyright{acmlicensed}\acmConference[WSDM '22]{Proceedings of the Fifteenth ACM International Conference on Web Search and Data Mining}{February 21--25, 2022}{Tempe, AZ, USA}
\acmBooktitle{Proceedings of the Fifteenth ACM International Conference on Web Search and Data Mining (WSDM '22), February 21--25, 2022, Tempe, AZ, USA}
\acmPrice{15.00}
\acmDOI{10.1145/3488560.3498532}
\acmISBN{978-1-4503-9132-0/22/02}

\title{Query Interpretations from Entity-Linked Segmentations}

\def\bsauthor{%
\vspace*{-1.5ex}%
Vaibhav Kasturia$^{*}$\qquad\quad
Marcel Gohsen$^{\dagger}$\qquad\quad
Matthias Hagen$^{*}$\\[1ex]
\gdef\bsauthor{%
Vaibhav Kasturia,
Marcel Gohsen, and
Matthias Hagen}}

\author{\bsauthor}
\affiliation{%
\institution{%
\parbox[t]{0.355\textwidth}{\centering
\mbox{\kern-0.5em $^{*}$Martin-Luther-Universit{\"a}t Halle-Wittenberg}\\
$<$first$>$.$<$last$>$@informatik.uni-halle.de}%
\qquad%
\parbox[t]{0.3\textwidth}{\centering
\mbox{\kern-0.5em$^{\dagger}$Bauhaus-Universit{\"a}t Weimar}\\
marcel.gohsen@uni-weimar.de}
\vspace{0.5ex}}
\city{}
\country{}
}

\begin{abstract}
Web search queries can be ambiguous: is \query{source of the nile} meant to find information on the actual river or on a board game of that name? We tackle this problem by deriving entity-based query interpretations: given some query, the task is to derive all reasonable ways of linking suitable parts of the query to semantically compatible entities in a background knowledge base. Our suggested approach focuses on effectiveness but also on efficiency since web search response times should not exceed some hundreds of milliseconds. In our approach, we use query segmentation as a pre-processing step that finds promising segment-based ``interpretation skeletons''. The individual segments from these skeletons are then linked to entities from a knowledge base and the reasonable combinations are ranked in a final step. An experimental comparison on a combined corpus of all existing query entity linking datasets shows our approach to have a better interpretation accuracy at a better run time than the previously most effective methods.
\end{abstract}



\maketitle

\section{Introduction} \label{sec:introduction}

We deal with the task of entity-based query interpretation: given a keyword query, automatically identify the most reasonable interpretation(s) based on the potentially mentioned (named) entities from a knowledge base. Web search log studies have shown that more than 70\%~of the queries contain entities~\cite{guo:2009} and that more than 50\%~of the queries directly refer to an entity or a set of entities~\cite{pound:2010, mika:2013}. For many such queries, identifying the actually meant entities helps to resolve ambiguities and thus to show better search results~\cite{anagnostopoulos:2005,chaudhuri:2007,mishra:2011} but also to populate the entity cards / knowledge box next to the search results with appropriate entries~\cite{hasibi:2017}.

In the field of natural language processing~(NLP), entity recognition aims to identify spans of words that belong to some entity class but linking them to a concrete entity is not required (e.g., recognizing ``Kim'' as a person suffices---mapping it to a concrete instance is not necessary). And mostly run on well-formed text, many entity recognition approaches make use of part-of-speech tags usually obtained by deriving parse trees---a rather error prone process for ungrammatical texts like keyword queries. Thus, instead of employing some NLP~entity \emph{recognition} approach, query understanding methods from the IR~field ``directly'' try to \emph{link} (consecutive) query terms to entities in some given background knowledge base~\cite{balog:2018}.

For ambiguous queries like \query{paris hilton} (celebrity vs.\ a hotel in Paris), linking query terms to entities helps to recognize the different interpretations and thus to disambiguate or to potentially diversify the knowledge box entries and search results. The ``only'' problem is to come up with the celebrity, the hotel group, and the city as mentioned entities and to figure out that from the more than 20~cities named Paris probably the capital of France is meant (none of the other Parises has a Hilton hotel). At the same time, the whole procedure needs to be very fast to not increase the web search response time. For this reason, many entity linkers only target the most prominent entities such that often only one single query interpretation can be inferred (probably the celebrity in the \query{paris hilton} example)---missing the chance of diversifying the results and knowledge boxes in case of two or more reasonable interpretations. Our new approach instead can return multiple interpretations.

To formalize the problem of entity-based query interpretation, we distinguish between three types of entities that were not clearly separated in previous query--entity studies: explicit, implicit, and related entities. Explicitly mentioned entities appear in some ``standard'' surface form (e.g., \query{paris}) while implicitly mentioned entities appear as a ``description'' (e.g., \query{capital of france}). Identifying the most likely combinations of explicit (and possibly implicit) entities in a query is the basis of query interpretation---implicit entities often even form (part of) an ``answer''. In contrast, related entities are not part of an interpretation but rather retrieved to further populate the knowledge box (e.g., showing the Eiffel tower alongside Paris).

Many query--entity linkers follow a ``dumb'' brute-force approach: checking for every combination of candidate entities whether they form a meaningful query interpretation (often even including related entities that cannot be part of an interpretation). Instead, we ``inform'' the combination phase by a query segmentation pre-processing. Query segmentation is the task of finding consecutive query terms that ``belong together'': the query segments. Our approach runs a fast query segmenter to identify the most promising segmentations and then uses them as ``skeletons'' for the combination phase that only has to link the few contained segments to explicit and implicit entities. The best combinations of linked and unlinked segments are then ranked as possible interpretations at a better run time than that of brute-force entity linkers.

Our contributions are threefold: \Ni we are the first to use query segmentation as an entity linking subroutine to develop a very fast approach that is able to find more than one query interpretation by combining sets of ``compatible'' entity mentions, \Nii we carefully merge, re-annotate, and extend the publicly available query--entity corpora to create a new corpus of 2,800~queries with interpretations, and \Niii we conduct a large-scale comparison of existing entity linking and query interpretation approaches on the new corpus.

The comparison shows our new approach to have a better interpretation accuracy at a much better run time than the previously most effective method (47\,ms~per query instead of~282\,ms). Our new corpus and our approach's implementation are publicly available.\footnote{\url{https://webis.de/data/webis-qinc-22.html}, \url{https://github.com/webis-de/WSDM-22}}

\section{Related~Work} \label{sec:related-work}
We review approaches and datasets for query entity linking or disambiguation, for query interpretation, and for query segmentation.

\subsection{Entity Linking}
Entity linking methods for queries (or short texts in general) can be categorized as \emph{explicit}, \emph{implicit}, or \emph{related} entity linking / retrieval. Explicit entities appear in a ``standard'' surface form (e.g., \query{paris}), implicit entities appear as a ``description'' (e.g., \query{capital of france}), while related entities are not part of the query (e.g., the Eiffel tower).

\paragraph{Explicit Entity Linking.}
TAGME~\cite{ferragina:2010} as one of the first entity linkers for short texts scores candidate entities by their commonness as Wikipedia anchor texts and the likelihood to co-occur with other candidates. Later variants improved the mention detection~\cite{piccinno:2014,cornolti:2016}---at the cost of a time-consuming retrieval phase. The fast FEL~entity linker of \citet{blanco:2015} combines Wikipedia anchor texts with log information about entities in clicked search results. Hashing and compression techniques reduce the system's memory footprint of pre-computed commonness scores and word embeddings. \citet{hasibi:2015,hasibi:2016,hasibi:2017b} developed Nordlys---its most effective entity linker combines commonness with language models~\cite{ogilvie:2003} to detect and rank candidates; later extended to also identify entity types~\cite{garigliotti:2019}. 

\paragraph{Implicit Entity Linking.}
\citet{perera:2016} were the first to explicitly target the detection of implicit entities---for tweets. They combine knowledge graphs and temporal context to identify explicit entities popular at the time of a tweet. Recently, \citet{hosseini:2021} suggested a respective learning-to-rank approach.

\paragraph{Related Entity Retrieval.}
Nordlys~\cite{hasibi:2015,hasibi:2016,hasibi:2017b} also offers an entity retrieval routine that somewhat mixes explicit, implicit, and related entities. \citet{bi:2015} and \citet{huang:2018} propose methods that particularly target related entities for a query while \citet{dietz:2019} later proposed an approach to rank entities related to a given entity.

\paragraph{Discussion.}
Most of the above approaches focus on precision: they try to find the most likely entity per mention. But for query interpretations (i.e., reasonable combinations of linked entities) also other entities may be important. We thus treat entity linking as a recall-oriented pre-processing. A subsequent combination phase will then use query segmentation skeletons to achieve convincing recall and precision scores for the actual query interpretations. These interpretations  are the main focus of our approach; entity linking is an important but ``just'' one of the subroutines.

\subsection{Entity Disambiguation}
Nowadays, entity linking and disambiguation are often used synonymously~\cite{rizzo:2017} but originally, linking meant the recall-oriented recognition of candidates for a mention, while disambiguation referred to the precision-oriented selection of the most suitable entity.

\citet{cucerzan:2007} tackled entity disambiguation by maximizing the ``global'' similarity of the disambiguations of all mentions, while \citet{ratinov:2011} used a previous ``local'' step to find the top entity per mention. \citet{hoffart:2011} suggest to find the densest knowledge subgraph that contains all mention nodes and exactly one mention--entity edge per mention. Their approach was improved by \citet{nguyen:2014} who first disambiguate low ambiguity mentions and then use these to disambiguate the rest. With the rise of neural approaches, \citet{yamada:2016} suggested to combine word--entity embeddings with graph-based approaches in a learning-to-rank setup. Also the RadboudEL tool~\cite{vanhulst:2020} uses embeddings to disambiguate mentions detected with Flair NER~\cite{akbik:2018}.

\paragraph{Discussion.} Query interpretation is a disambiguation problem: among the possible combinations of entity-linked mentions in a query, the most suitable have to be chosen based on ``disambiguating'' the mentions with more than one entity link. For analyzing the semantic coherence of a query interpretation, our own approach adapts and combines ideas from the disambiguation literature.

\subsection{Query Interpretation}

Entity-based query interpretations help to populate the knowledge box next to the search results and to retrieve better results than with ``simple'' term matching---demonstrated by entity-informed learning-to-rank~\cite{xiong:2015,xiong:2016,xiong:2017} and language model-based retrieval~\cite{raviv:2016}. 

\citet{sawant:2013} suggest to ``interpret'' queries over structured knowledge bases from users who are not aware of the underlying schema (e.g., the query \query{parents paris hilton} may not work but two separate queries for her mother and father). This setting is related to our notion of implicit entities but substantially differs from our notion of the query interpretation task that first was prominently featured in the short-text track of the ERD~Challenge~\cite{carmel:2014}: generating the possibly multiple combinations of non-overlapping entity mentions that are ``semantically compatible''. Still, many approaches submitted to the challenge identify only one interpretation per query---among them the top-2 systems~\cite{cornolti:2016,chiu:2014}. 

After the ERD~Challenge, \citet{hasibi:2015,hasibi:2016,hasibi:2017b} added a so-called greedy interpretation finding~(GIF) method to their Nordlys system to generate multiple interpretations. Still, in their experiments, the GIF~method also often did find only one interpretation. 

\subsection{Query--Entity Datasets}\label{sec:query-corpora}

Three publicly available datasets of queries annotated with entities are used in the entity linking / query interpretation literature. \citet{balog:2013} introduced the ES-DBpedia dataset with 485~queries that was later revised by \citet{hasibi:2017c} (DBpedia-Entity~v2, 467~queries). The newer version contains relevance judgments for the entities (0:~irrelevant, 1:~relevant, 2:~highly relevant) but since crowdsourcing was used, some judgments are a bit ``noisy'' (e.g., Julia Lennon as the mother of John Lennon is highly relevant for the query \query{john lennon parents} while the father Alfred Lennon is only relevant without the fatherhood being controversial).

The ERD~Challenge only released 91~queries for training while the YSQLE dataset~\cite{Yahoo:YSQLE} as the largest publicly available query--entity corpus has 2,635~queries with Wikipedia entities manually labeled as ``true'' if part of some user intent and ``false'' otherwise.  

\paragraph{Discussion.} Already~\citet{erp:2016} have noted that different characteristics and annotation schemes make it difficult to compare entity linking results for different datasets. Furthermore, the DBpedia-Entity~v2 and YSQLE~datasets actually do not contain annotated query interpretations (only entity-level labels of ``relevance'' in case of DBpedia-Entity~v2 and only an indication whether an entity is part of an interpretation but not which one if there were more than one in case of YSQLE). To form one large and homogeneous dataset, we thus decided to combine and re-annotate all three query--entity datasets with interpretations. We (re-)evaluate the known entity linking and query interpretation approaches on our new dataset to ensure comparability with previous results.

\subsection{Query Segmentation} \label{sec:related-work-query-segmentation}
Interestingly, so far, query interpretation and entity linking approaches ignore the related task of query segmentation (identifying consecutive query terms that ``belong'' together---but without linking them to a knowledge base). For the first time, we suggest to integrate segmentation and entity linking. In our approach, query segmentation forms a pre-processing step to identify promising ``skeletons'' that are then entity-linked to derive interpretations.

We use the fast approaches of \citet{stein:2011e,stein:2012q} since they are easy to implement and do not rely on large query logs like some other methods~\cite{mishra:2011,li:2011}, and since they achieved a better segmentation accuracy than earlier unsupervised approaches~\cite{risvik:2003, jones:2006, huang:2010} at a good impact on retrieval effectiveness~\cite{roy:2012,stein:2012q}. We decided against the approach of \citet{wu:2015} that equips the methods of \citet{stein:2011e,stein:2012q} with a post-processing step since in pilot experiments the run time requirements of our re-implementation were much higher than the only 1--2\,ms~per query of the simpler methods.

\section{Problem Definition} \label{sec:problem-definition}

We view a query~$q$ as a sequence~$t_1, t_2, \ldots, t_n$ of terms. A segment~$s$ of~$q$ is a contiguous term subsequence (e.g., \query{neil moon} is not a segment of the query \query{neil armstrong moon}) and a valid segmentation~$S$ for~$q$ consists of disjunct segments~$s$ whose concatenation is~$q$. Query segmentation aims to find the ``best'' segmentation(s) for a query in the sense of retrieval effectiveness (treating segments as phrases to be matched in search results). But we are ``only'' interested in the skeletons of segment boundaries to try to link the segments to entities and to form entity-based query interpretations.

Named entities are uniquely identifiable real or fictional objects often organized in taxonomies (we use 108~classes from the taxonomy of \citet{Sekine:2002}) and stored in knowledge bases (we use Wikipedia but any other knowledge base is possible). A segment~$s$ is a \emph{mention} if it refers to an entity~$e$ (e.g., the segment \query{neil armstrong} is a mention of \query{Neil\_Armstrong}, the astronaut). A \emph{surface form} is a mention that is commonly used as a reference for some entity.

We differentiate three types of entities in a query: explicit, implicit, and related entities. \emph{Explicit entities} are mentioned by their actual name or a surface form (e.g., \query{neil armstrong}) but they may overlap like in the query \query{elton john wayne rooney} with the explicit entities \query{Elton\_John}, \query{John\_Wayne}, and \query{Wayne\_Rooney}.

\emph{Implicit entities} have mentions that ``describe'' them (e.g., the query \query{first man on the moon} has the implicit entity \query{Neil\_Armstrong}). The \emph{span} of an implicit entity is the shortest query segment that accurately describes the entity without causing misunderstandings (e.g., \query{first man} alone would be ambiguous). Implicit entities occur when a ``modifier''~\cite{imrattanatrai:2018} is added to an entity or concept (e.g., \query{first man on the} before the entity \query{Moon} (surface form of the astronomical body orbiting Earth) leads to the implicit entity \query{Neil\_Armstrong}). By removing the term \query{first}, the query \query{man on the moon} then currently has twelve astronauts as implicit entities---besides several books, movies, and pieces of music as explicit entities. Note that even when an implicit entity does not span the whole query, it could be the answer (e.g., \query{Neil\_Armstrong} answers \query{who~was~the~first man on the moon} even though only the last five terms are the span).

Special forms of implicit entities are \emph{Vossian antonomasias}~\cite{fischer:2020} with explicit entities as modifiers (e.g., \query{mozart of chess} referring to \query{Magnus\_Carlsen}), and nestings of modifiers like \query{birthplace of first man on the moon} that require to find the inner implicit entity \query{Neil\_Armstrong} to get to the outer implicit entity \query{Wapakoneta,\_Ohio} (span of the outer entity containing the span of the inner entity). 

\emph{Related entities} are not explicitly or implicitly mentioned in a query but have a semantic relation to an explicit or implicit entity (e.g., \query{Buzz\_Aldrin} is a related entity for the query \query{neil armstrong}). While explicit and implicit entities help to understand a query's intent and to populate the knowledge box next to the search results, related entities are mainly useful for the knowledge box.

Combining query segmentation and query--entity linking, we view an \emph{entity-based query interpretation} as a valid query segmentation in which mentions are linked to entities. To form a plausible interpretation, the linked and the unlinked segments must be semantically compatible. For example, in the query \query{armstrong and its surroundings}, the segment \query{armstrong} clearly is not a mention of \query{Neil\_Armstrong} since the terms \query{its surroundings} clarify that some of the many places named Armstrong is meant but not a person.

Two interpretations of a query (different segmentations or linked entities) are \emph{equivalent} iff they express the same information need. For example, the query \query{apollo 11 mission duration} has two equivalent interpretations that ask for the time span of about 8~days: \interpretation{Apollo\_11\,|\,mission duration} and \interpretation{Apollo\_11\,|\,duration} (the former links \query{apollo 11} to \query{Apollo\_11} and has \query{mission duration} as a second segment while the latter links \query{apollo 11 mission} to \query{Apollo\_11}). Still, the notion of ``same information need'' is rather fine-granular (e.g., the query \query{blue line schedule} has the non-equivalent interpretations \interpretation{Blue\_Line\_(Delhi\_Metro)\,|\,schedule} for Delhi-based users and \interpretation{Blue\_Line\_(Montreal\_Metro)\,|\,schedule} for Montreal-based users that on a coarse-grain level both ask for a metro schedule).

A bit tricky are queries that mention the same entity explicitly and implicitly like the fact-checking question \query{was neil armstrong the first man on the moon}. A plausible interpretation could be to just map \query{neil armstrong} to \query{Neil\_Armstrong} and leave the segment \query{first man on the moon} ``unlinked'' but also linking both to the entity can make sense to directly come up with the yes-answer.

Entity-based query interpretations can help a search engine to understand the user intent, to show entities in a knowledge box, or to even potentially diversify the search results and knowledge box in case of more than one non-equivalent interpretation. In its essence, entity-based query interpretation is a query--entity linking task that aims for finding plausible combinations of linked entities and non-linked context segments. While in the query--entity literature often only the precision / recall of the individually linked entities is measured, the task of entity-based query interpretation is more ``strict'' since all entity links have to be detected in order for an interpretation to be correct. Note that, somewhat by definition, query segmentation is a natural pre-processing step of entity-based query interpretation. We demonstrate that this pre-processing leads to efficiency gains over traditional entity linking approaches while at the same time achieving better interpretation effectiveness (i.e., better combined linking / disambiguation effectiveness to identify more of the semantically reasonable mention combinations).

\section{Query Interpretation Approach} \label{sec:approach}
Most query interpretation methods follow a two-step schema. \Ni In a candidate linking phase, they identify potential entities from a background knowledge base for each of the $O(n^2)$~segments of an $n$-term query. In this process, often also many related entities are found as potential links instead of just explicit and implicit entities. \Nii In a rather time-consuming combination and ranking phase, the methods then try to identify meaningful combinations of the  many candidate links per segment and rank the combinations by scoring how well the linked entities ``fit together''. Interestingly, many methods finally still output only one interpretation even though many queries may have more than one reasonable interpretation (cf.~Section~\ref{sec:corpus} for a respective analysis of our new corpus).

In our approach, we reduce the run time of the costly combination phase. To this end, we run a query segmentation pre-processing phase parallel to the candidate linking phase. In pilot experiments, we tried to first run segmentation and then only forward the segments from the most promising segmentations to the candidate phase. But running segmentation and brute-force linking of all $O(n^2)$~segments in parallel was slightly faster than a sequential processing. Thus, the most promising segmentations are not used to reduce the candidate effort but as the ``skeletons'' to be filled in the combination phase. This often vastly reduces the search space of the combination phase and overall saves run time.

\subsection{Query Segmentation Phase} \label{sec:approach-segmentation-phase}
In a first step, our entity-based query interpretation approach identifies promising segmentations. Query segmentation methods aim to rank the possible $2^n$~valid segmentations of an $n$-term query according to retrieval effectiveness when the segments were treated as phrases to be matched in search results. Segmentation is a query understanding step---but without actually ``asking'' which of the segments are entities and which are just (common) phrases. We use segmentations to reduce the run time of the combination phase that ``only'' will need to fill the most promising segmentation skeletons.

In pilot experiments, we re-implemented and evaluated the best unsupervised query segmentation methods from the literature that do not need query logs~\cite{stein:2011e,stein:2012q,wu:2015}. From these, the slightly more effective approach of~\citet{wu:2015} was slower than the simpler approaches of~\citet{stein:2011e,stein:2012q} such that we decided to employ the simpler approaches in our actual system (cf.~Section~\ref{sec:optimizing-segmentation-phase} for more details on the setup). The approaches of~\citet{stein:2011e,stein:2012q} rank the possible segmentations of a query by summing up pre-computed segment weights stored in a hash table for quick access. The difference between the approaches is the weight pre-computation (some approaches just assign weights to phrases that are strict noun phrases or titles of Wikipedia articles, etc.).

\begin{table}[t]
  \caption{Segmentations of \query{new york times square dance} with the scores of~\citet{stein:2011e}. Lower ranked segmentations with the same highest-weight segment in gray (our filter heuristic removes them). The column ``Ratio'' indicates the ratio of the scores of two consecutive non-gray segmentations.} 
  \label{table-segmentation-example}
  \begin{footnotesize}
    \begin{tabular}{@{}rlrr@{}}
      \toprule
      \addlinespace[2pt]
        Rank & \multicolumn{1}{c}{Segmentation~$S$}    & $\mathit{score}(S)$ & Ratio \\
      \midrule
           1 & \tt new york times | square dance       &       496.6 million &       \\
      \textcolor{lightgray}{2} & \textcolor{lightgray}{\tt new york times | square | dance}     &       \textcolor{lightgray}{496.2 million} \\
           3 & \tt new york | times square | dance     &       333.4 million & 0.671 \\
      \textcolor{lightgray}{4} & \textcolor{lightgray}{\tt new york | times | square dance}     &       \textcolor{lightgray}{331.0 million} \\
      \textcolor{lightgray}{5} & \textcolor{lightgray}{\tt new york | times square dance}       &       \textcolor{lightgray}{330.8 million} \\
      \textcolor{lightgray}{6} & \textcolor{lightgray}{\tt new york | times | square | dance}   &       \textcolor{lightgray}{330.8 million} \\
           7 & \tt new | york times | square dance     &        35.6 million & 0.107 \\[-1ex]
      \vdots & \multicolumn{1}{c}{\vdots}              &              \vdots \\
          13 & \tt new | york | times square dance     &                 312 \\
          14 & \tt new | york | times | square | dance &                   0 \\
          15 & \tt new | york times square dance       &                  -1 \\
          16 & \tt new york times square dance         &                  -1 \\
      \bottomrule
    \end{tabular}
  \end{footnotesize}
\end{table}

Table~\ref{table-segmentation-example} shows how the approach of~\citet{stein:2011e} processes the query \query{new york times square dance}. The score of a segmentation~$S$ is the sum of the contained segments' weights. In case that $s$~is not a title or re-direct of some Wikipedia article, the weight of~$s$ simply is its occurrence frequency in the Google n-gram corpus~\cite{brants:2006} multiplied by the length of~$s$ in words. In case that $s$~is a title or re-direct of some Wikipedia article, the weight of~$s$ is (1 + occurrence frequency of the most frequent word-2-gram contained in~$s$) multiplied by the length of~$s$ in words. For example, the 3-gram \query{new york times} gets as its weight the occurrence frequency of \query{new york} + 1 (i.e., 165.4~million + 1) multiplied by~3 for a total weight of 496.2~million. The final score of 496.6~million for the segmentation \query{new york times\,|\,square dance} is the sum of this 496.2~million weight for \query{new york times} and the 420,880~weight of \query{square dance} (2 times the 210,440~frequency of \query{square dance} that actually also is a Wikipedia article title but also its own most frequent sub-word-2-gram). Single-term segments do not contribute to the score of a segmentation in the approach of~\citet{stein:2011e} such that the ``non-segmentation'' \query{new\,|\,york\,|\,times\,|\,square\,|\,dance} gets a score of~0. Special cases are segmentations with segments that do not have a frequency in the Google n-grams; such segmentations are assigned a score of~$-1$ to be ranked below the non-segmentation.

The top-ranked segmentation \query{new york times\,|\,square dance} is not entity-linked but already hints at an interpretation of searching articles in \query{The\_New\_York\_Times} about the folk dance \query{Square\_Dance}, the \query{Square\_Dance\_(ballet)}, or the \query{Square\_Dance\_(film)}, while the third-ranked segmentation \query{new york\,|\,times square\,|\,dance} hints at a search for \query{dance} events on the \query{Times\_Square} in \query{New\_York\_City}.

Most of the $2^n$~segmentations of an $n$-term query do not yield plausible interpretations. On our manually annotated training set (cf.~Section~\ref{sec:train-test}), a pilot study showed that often only the highest-scoring segmentation that includes some particular segment is used as an interpretation skeleton and that lower-ranked segmentations with huge score differences to higher-ranked segmentations are hardly used as skeletons. We thus apply respective filter heuristics to not forward all segmentations to the combination phase. A first filter removes segmentations whose highest weighting segment is contained in a higher-ranked segmentation. In the example of Table~\ref{table-segmentation-example}, the potentially interesting segmentation \query{new york\,|\,times\,|\,square dance} (find times of square dance events) is removed since \query{new york} also is included in the third-ranked segmentation. A second filter removes segmentations for which the score ratio to the lowest kept higher-ranked segmentation falls below some threshold like $0.66$ or $0.5$ (threshold trained, cf.~Section~\ref{sec:optimizing-segmentation-phase}). In the example, the seventh-ranked segmentation and all below are removed.

\subsection{Candidate Linking Phase} \label{sec:approach-linking-phase}
In parallel to the query segmentation, the candidate linking tries to find entities for all the $O(n^2)$~potential segments. Ideally, the linking would identify explicit and implicit entities. Practically, this depends on the knowledge base and the potentially contained implicit mentions (e.g., the Wikipedia knowledge base that we use has a redirect that actually maps \query{first man on the moon} to \query{Neil\_Armstrong}). Going beyond the explicit and implicit aliases of the used knowledge base is beyond the scope of this paper. However, including a real implicit entity linker (e.g., for Vossian antonomosias) is an interesting task for future work. Our new corpus (cf.~Section~\ref{sec:corpus}) could be a good starting point, since we also have annotated the implicit entities. 

Our entity linking module is based on titles of Wikipedia articles, redirects, and disambiguation pages (in practice, of course, also any other knowledge base can be used). The about 13~million distinct key--value pairs (keys are potential query segments and values are lists of entities that can be referred to by this segment) are stored in a RocksDB table~\cite{facebook:rocksdb} for fast exact-match access and in a Lucene index~\cite{apache:lucene} to quickly find imperfect matches (in a pilot study, a depth of 150~Lucene results was best). The exact and imperfect matches for segmentations selected in the query segmentation phase are passed to the combination phase. For the example from Table~\ref{table-segmentation-example}, the segment \query{new york times} is linked, among others, to the entity \query{The\_New\_York\_Times} and \query{square dance} is linked to \query{Square\_Dance}.

\subsection{Combination and Ranking Phase} \label{sec:approach-combination-phase}
To derive entity-based query interpretations, our approach computes \emph{commonness} scores for mentions in the promising segmentations. Commonness (i.e., the likelihood of an entity--mention link) is used in many linkers as a solid baseline~\cite{ji:2011}. We use the same Wikipedia dump that forms our background knowledge base and apply the commonness computation of \citet{ferragina:2010}.

To ``fill'' a skeleton forwarded by the segmentation phase, we order the potential entities for each segment by their commonness score (discarding entities with a 0-commonness) and as a fall back solution add the option of not linking a segment to any entity but to keep it as a phrase (i.e., put in quotes in a web search). The potential interpretations can then be derived by a Cartesian product of the not-0-common entities and the unlinked respective segments. In the above example, the top-segmentation \query{new york times\,|\,square dance} has eight interpretation candidates (\query{The\_New\_York\_Times} or no link for the segment \query{new york times}, and the folk dance, the ballet, the film, or nothing for the segment \query{square dance}).

In the interpretation ranking, we combine three weights from the entity linking literature: \Ni the above described commonness~$\mathrm{CMN}$, \Nii the likelihood of two entities to occur together (relatedness~$\mathrm{REL}$), and \Niii the likelihood of an entity to occur with the unlinked segments (context~$\mathrm{CXT}$). We compute the relatedness and context weights using Wikipedia-based joint word--entity embeddings~\cite{wikipedia2vec:wikipedia2vec} provided by \citet{yamada:2016} (data from April~2018 but matching our background Wikipedia knowledge base). We use the configuration suggested by the authors since it also performed best in our pilot experiments: average cosine similarity of an entity's embedding vector with the other entities in an interpretation (relatedness) or with the unlinked segments in an interpretation (context).

An interpretation~$I$'s score is the averaged weighted sum of the commonness, relatedness, and context scores of the entities~$e \in I$:

\begin{equation*}
\mathit{score}(I) =  \frac{1}{|\{e \in I\}|} \cdot \sum_{e \in I} \left(\alpha \cdot \mathrm{CMN}(e) + \beta \cdot \mathrm{REL}(e) + \gamma \cdot \mathrm{CXT}(e)\right),
\end{equation*}

with [0,1]-valued parameters $\alpha, \beta,$ and $\gamma$---often being the first idea to combine different weights, a linear combination also worked well in our pilot experiments. To optimize the parameters, we use hill climbing and maximize the interpretations' F\textsubscript{1}-score on our training set (cf.~Section~\ref{sec:train-test}). Interestingly, the configuration that best fits the training data is $\alpha= \beta = \gamma = 1$ such that commonness, relatedness, and context are of equal importance. An interpretation without linked entities has a 0-relatedness and will usually only be chosen when no entities have been found in the linking phase.
\section{Query Interpretation Corpus} \label{sec:corpus}
We combine, re-annotate, and extend the existing query--entity corpora to a coherent dataset for entity-based query interpretation: our new Webis Query Interpretation Corpus~2022 (Webis-QInC-22).

\subsection{Corpus Creation}

The available query--entity linking datasets~\cite{balog:2013,hasibi:2017b,carmel:2014,Yahoo:YSQLE} contain a total of 3,193~queries (mostly without annotated interpretations). After a normalization preprocessing (lowercasing, manual spell correction, special character normalization, etc.), the combined set consists of 2,598~unique queries to which we added 202~new queries that we found as ambiguity examples in different sources (e.g., \query{new york times square dance} from the query segmentation literature).

\paragraph{Entity and Interpretation Annotation}
To ensure a consistent annotation, a single main expert annotator (re-)annotated all the queries in the new corpus from scratch following guidelines adapted and extended from \citet{hasibi:2015}: \Ni an entity has to be an instance of the 108~entity classes from the taxonomy of \citet{Sekine:2002}, \Nii only explicit and implicit entities should be annotated (no related entities), \Niii the mention span of an entity is the shortest query segment that accurately refers to the entity (long enough to avoid misunderstandings), \Niv entities with overlapping mention spans are allowed in the sets of implicit and explicit entities but not as part of the same interpretation, \Nv entities in an interpretation have to be semantically and grammatically compatible with the other entities / segments in the interpretation, and \Nvi implicit entities are only part of an interpretation when they are not themselves the answer to the query. As the knowledge base for the annotation, we use the whole Web. Thus, mentions without a Wikipedia article may be linked to alternate resources (e.g., LinkedIn for ``ordinary'' people or Yelp for local companies). Entities without an easily assignable web resource have a comment added to the mention.

After an introduction to these guidelines, we performed a kappa test on a 50~query sample with two other annotators. This test indicated a high inter-annotator agreement for all annotation tasks (kappa scores of~0.65--0.7; usually termed as ``good'' or ``very good'' depending on the scheme). We used the few cases with disagreements in a discussion with the annotators to further fine-tune their understanding of the task. The main annotator then annotated all the queries in our corpus without access to the queries' previous annotations. To ensure a high quality, all the annotations were then reviewed by the two other annotators. Possible conflicts were discussed and adjusted until an agreement was reached.

The annotation process did result in at least one explicit or implicit entity being annotated for 2,234~queries. We then also checked the 163~entities contained in the previous corpora that had non-zero weights in their corpus but that were not selected by our annotators even after discussions. We kept 141~of them as related entities even though we did not (yet) annotate related entities for all queries in our new corpus since this is not the scope of our paper. Our annotator(s) also annotated relevance levels for the entities (1: might be part of a query intent, and 2: very likely is part of a query intent). Different to, for instance, the DBpedia-Entity v2~dataset that has more than 27,374~entities annotated with 0-scores (not relevant) for the 467~queries, we do not have 0-valued entities in our corpus but only include entities that might be part of a query interpretation.

To form interpretations, our annotator(s) were instructed to keep common phrases or concepts as segments and to indicate equivalent interpretations. Additionally, the interpretations were graded as ``plausible'' (1, few searchers might mean this interpretation), ``moderately likely'' (2, more searchers might mean this interpretation), or ``very likely'' (3, most searchers' intent would be this interpretation), and clarifying comments could be added (e.g., explaining why a~1 was assigned, etc.). In our running example, the interpretations \interpretation{The\_New\_York\_Times | Square\_Dance} and \interpretation{New\_York\_City\,|\,Times\_Square\,|\,dance} both got a score of~2 since among the good interpretations none is dominant.

\paragraph{Difficulty Assessment}
During and after the annotation, our annotators also assigned values on a 5-point scale for the annotation difficulty of a query based on the complexity / ambiguity of the mentioned entities. Easy queries with low ambiguity, few or no explicit entities, and no implicit entities get a value of~1 (e.g., the query \query{frank~zappa} only has the explicit entity \query{Frank\_Zappa}). Slightly more difficult queries with moderate ambiguity, potentially various explicit entities, or easy implicit entities get a value of~2 (e.g., the query \query{frank~zappa~mother} contains the explicit entity \query{Frank\_Zappa}, has the implicit entity \query{Rose\_Marie\_Zappa}, and possibly the implicit entity \query{The\_Mothers\_of\_Invention}, the band of \query{Frank\_Zappa}). Difficult queries with high ambiguity or complex implicit entities get a value of~3 (e.g., \query{judges fisa court 2005} with several implicit judge entities). Queries for which it is not clear what the user wanted get a value of~4 (e.g., for the query \query{windows xp 8}, does the user ask for an upgrade from \query{Windows\_XP} to \query{Windows\_8} or is the term~\query{8} just referring to something else?). Finally, queries with more than 20~entities get a value of~5 (e.g., the query \query{free online games} has plenty of implicit entities like \query{League\_of\_Legends}, \query{Dota\_2}, etc.)---when there were too many explicit or implicit entities, our annotators tried to add a link to a Wikipedia disambiguation page or a Wikipedia list. Overall, the average query difficulty is~1.77 in our dataset.            

\paragraph{Query Classes}
In a final annotation step, we broadly categorized the queries into five classes. \Ni \emph{Categorical} queries refer to a group of (often related) implicit entities (e.g., \query{members of u2}). \Nii \emph{Conceptual} queries come in two flavors: queries not containing any entities and requesting information about general concepts (e.g., noun phrases like \query{black powder ammunition}) or queries asking for concepts related to an explicit entity (e.g., \query{churchill downs horse racing track schedule} asking for a schedule (concept) of horse races taking place at the \query{Churchill\_Downs\_(racetrack)}). \Niii \emph{Question} queries are formulated as a question (e.g., \query{how~do sunspots~affect~us}). \Niv \emph{Relational} queries are similar to conceptual queries with the difference that the requested information for an explicit entity is an implicit entity and not a concept (e.g., \query{niagara falls origin lake} where \query{Niagara\_Falls} is the explicit entity and \query{Lake\_Erie} is the implicit entity with the entire query as mention span; also all Vossian antonomasias fall in this category). \Nv \emph{Surface} queries have a query string that is an explicit entity mention (e.g., \query{frank zappa}).

\paragraph{Interoperability} A study by \citet{erp:2016} somewhat criticized that different entity linking datasets often use just a single knowledge base such that a re-mapping might be needed in case of another preference. To ease using our dataset, we also include DBpedia~\cite{auer:2007}, Freebase~\cite{bollacker:2008}, Wikidata~\cite{vrandecic:2012}, and YAGO3~\cite{mahdisoltani:2015} links. 

\subsection{Corpus Analysis}
The queries in our Webis-QInC-22 on average have 3.62~terms (from~1 to~14), contain 1.99~explicit entities (0~to~19), and have 2.22~interpretations (1~to~40). Table~\ref{table-corpus-characteristics} details the characteristics per query length. Shorter queries of up to four terms are more ambiguous---less context given---and thus have more interpretations than longer queries while the average number of mentions increases with the length of the query (more segments = more potential of mentions). Some queries have no mentions at all (e.g., just concepts included) but every query has at least one interpretation. Since our guidelines stated to only have implicit entities in an interpretation when they are not the answer, only~59 of the 6,222~interpretations contain an implicit entity (e.g., \interpretation{Neil\_Armstrong} is not(!) an interpretation of \query{first man on the moon} but of course annotated as an implicit entity). As for entity-based query interpretation, the implicit entities thus seem not to be too important. However, when it comes to actually answering the query (e.g., retrieving entries for the knowledge box), the implicit entities often play a major role. We thus already have them in our corpus but in this paper focus on forming interpretations from the entities found as titles or re-directs of Wikipedia articles. Identifying also all the implicit entities and populating the knowledge box with them is an interesting task for future research.

\begin{table}[t]
  \caption{Characteristics of our new Webis-QInC-22 dataset.}
  \label{table-corpus-characteristics} 
  \begin{footnotesize}   \setlength{\tabcolsep}{4.5pt}
    \begin{tabular}{@{}crrrrrrrrrr@{}}
       \toprule
       \multirow{2}{3em}{Query Length} & Count & \multicolumn{3}{l}{Mentions} & \multicolumn{3}{l}{Explicit Entities} & \multicolumn{3}{l}{Interpretations}\\ 
       \cmidrule(l){3-5} \cmidrule(l){6-8} \cmidrule(l){9-11} 
                                       &       & Min &  Avg & Max &                      Min &  Avg & Max &                    Min &  Avg & Max\\
       \midrule                                                                                                                
                                     1 &   206 &   0 & 0.86 &   1 &                        0 & 2.47 &  19 &                      1 & 2.79 &  19\\
                                     2 &   610 &   0 & 1.08 &   3 &                        0 & 2.16 &  19 &                      1 & 2.60 &  40\\
                                     3 &   775 &   0 & 1.17 &   4 &                        0 & 2.07 &  18 &                      1 & 2.40 &  40\\
                                     4 &   540 &   0 & 1.34 &   4 &                        0 & 2.00 &  19 &                      1 & 2.13 &  30\\
                                     5 &   290 &   0 & 1.51 &   5 &                        0 & 1.65 &   9 &                      1 & 1.60 &  16\\
                                     6 &   154 &   0 & 1.56 &   4 &                        0 & 1.74 &  11 &                      1 & 1.81 &  16\\
                                     7 &    96 &   0 & 1.76 &   5 &                        0 & 1.56 &   7 &                      1 & 1.47 &  16\\
                                 8--14 &   129 &   0 & 1.83 &   5 &                        0 & 1.38 &   7 &                      1 & 1.30 &   8\\
       \midrule                                                                                                                   
                                 1--14 & 2,800 &   0 & 1.27 &   5 &                        0 & 1.99 &  19 &                      1 & 2.22 &  40\\
       \bottomrule 
    \end{tabular}
  \end{footnotesize}
\end{table}

\section{Evaluation}\label{sec:evaluation}
After selecting the segmentation method, tuning the segmentation filtering threshold, and tuning the combination score parameters on a fixed training set, we evaluate the individual steps of the entity-based query interpretation process. Note that others might later choose different tuning ideas but can use the same train--test split. 

\subsection{Webis-QInC-22 Train--Test Split} \label{sec:train-test}
We want to ensure that highly similar queries (e.g., \query{lake murray} and \query{lake murray fishing}) are not separated in a train--test split such that no direct information on entities is leaked from training to test. To this end, we use the YSQLE~dataset's session information and manually combine other similar queries to ``clusters'' that should not be split. Since the split should also respect the query category and length distributions, we use a hill climbing optimization~\cite{wikipedia:hillclimbing} starting from a random cluster-respecting 80-20~split of the queries. The ``error'' of a split is the sum of the absolute differences between the train/test sets' distributions and the distributions in the whole corpus. As long as the error exceeds a given threshold, two random clusters are exchanged between train and test set. Our derived split has 2256~queries for training and 544~queries for testing. 

\subsection{Optimizing the Segmentation Phase} \label{sec:optimizing-segmentation-phase}
The segmentation phase ideally finds all skeletons of the ground-truth interpretations at a good precision (of the derived skeletons against the skeletons of the ground truth) since otherwise the combination phase is slowed down trying to fill too many skeletons. A segmentation is a \emph{complete match} when it exactly matches the skeleton of a ground-truth interpretation while a \emph{partial match} may sub-segment segments that are not mentions (e.g., the segmentation \query{new york times\,|\,square\,|\,dance} is no match for the interpretation \interpretation{The\_New\_York\_Times\,|\,Square\_Dance} (\query{square dance} is split) but a partial match for \interpretation{The\_New\_York\_Times\,|\,square\,dance} that does not link \query{square dance} to an entity). In other words, a partial match may get the segmentation wrong for non-linked ground-truth segments.

From the potential $2^n$~segmentations only those are passed to the combination phase that do not have the same highest weighting segment as a higher ranked segmentation and whose score ratio compared to the previous passed-on segmentation does not fall below a threshold (cf.~Section~\ref{sec:approach-segmentation-phase}). We optimize this threshold for the approaches of \citet{stein:2011e,stein:2012q} with respect to the $F_{1}$~score of completely / partially matched passed-on skeletons (the combination phase will only be able to fill these skeletons).

\begin{table}[t]
  \caption{Segmentation results of Hagen et al.'s approaches~\cite{stein:2011e,stein:2012q} on our train set. CSA/PSA, CSB/PSB: complete/partial skeleton matches on all or only the better interpretations (better = assessment of~2 or~3). Time: avg.\ time per query.}
  \label{table-segmentation-evaluation} 
  \begin{footnotesize}
    \begin{tabular}{@{}lrrrrrr@{}}
      \toprule
      Segmentation                  & \multicolumn{4}{l}{Recall}                                        && \multirow{2}{2em}{Time \null\;(ms)} \\ 
      \cmidrule(l){2-5}
                                    &           CSA  &           CSB  &           PSA  &           PSB  \\
      \midrule
      No Segmentation               &         0.226  &         0.208  &         0.541  &         0.528  &&                                0.00 \\[.75ex]
      Na{\"i}ve \cite{stein:2011e}  &         0.855  &         0.846  &         0.924  &         0.915  &&                                0.72 \\
      Wiki-based \cite{stein:2011e} &         0.858  &         0.849  &         0.922  &         0.914  &&                                1.01 \\
      WT \cite{stein:2012q}         & \textbf{0.927} & \textbf{0.920} & \textbf{0.956} & \textbf{0.949} &&                                0.76 \\
      WT+SNP \cite{stein:2012q}     &         0.883  &         0.874  &         0.935  &         0.927  &&                                6.02 \\
      \bottomrule 
    \end{tabular}
  \end{footnotesize}
\end{table}

Table~\ref{table-segmentation-evaluation} shows the results with the best thresholds on the training set and what a no-segmentation would achieve (i.e., just one segmentation with each query term its own segment). As our segmentation strategy, we choose the WT~approach (threshold of~0.66). It achieves the highest partial and complete recall (it favors titles and redirects of Wikipedia articles and thus matches our background Wikipedia knowledge base) and is one of the fastest approaches (e.g., the WT+SNP~approach needs time for POS-tagging the query).

\subsection{Entity Linking Comparison}
We compare our simple brute-force entity linking approach to existing linking and recognition tools. Since recognition approaches are not required to link detected mentions to an actual entity~\cite{rizzo:2017}, we count every correctly recognized mention as a match for them.

In our evaluation, we derive micro and macro averages for precision, recall, and weighted recall of a method's results, where the combined ``classes'' for the macro-averaging are the individual queries' sets of entities. For a query~$q$ with ground-truth entity set~$E$ (containing relevance levels~$\mathit{rel}(e)$ for entities~$e$), the micro precision~$\mathit{prec}(q)$, recall~$\mathit{rec}(q)$ and weighted recall~$\mathit{rec}^*(q)$ of an approach's derived entity set~$E'$ are defined as:

\begin{align*}
    \mathit{prec}(q) &=
    \begin{cases}
      \frac{|E \cap E'|}{|E'|}\,, & \text{if } |E'| > 0\,, \\
                       \quad 1\,, & \text{if } |E'| = 0, |E| = 0\,, \\
                       \quad 0\,, & \text{if } |E'| = 0, |E| > 0\,;
    \end{cases}\\[.5ex]
    \mathit{rec}(q) &=
    \begin{cases}
      \frac{|E \cap E'|}{|E|}\,, & \text{if } |E| > 0\,, \\
                      \quad 1\,, & \text{if } |E| = 0, |E'| = 0\,,\\
                      \quad 0\,, & \text{if } |E| = 0, |E'| > 0\,;
    \end{cases} \\[.5ex]
    \mathit{rec}^*(q) &= \frac{\sum_{e' \in E' \cap E} \mathit{rel}(e')}{\sum_{e \in E} \mathit{rel}(e)}\,.
\end{align*}

Table~\ref{table-entity-linking-evaluation} shows the entity linking results of the publicly available entity linking and recognition systems evaluated on the explicit entities of our corpus. Since the later combination and ranking phase of entity-based query interpretation can only use entities recalled in the linking phase, a high recall---achieved by our brute-force entity linker---is desirable to not miss any interesting entity. For completeness, we also include the precision scores for all entity linkers in Table~\ref{table-entity-linking-evaluation} even though the actual query interpretation results in Table~\ref{table-interpretation-evaluation} clearly show that a recall-oriented entity linking helps to find more of the interesting interpretations: supporting our idea of a recall-oriented candidate linking phase.

\begin{table}[t]
  \caption{Entity linking results on our test set in form of micro/macro~(Mic/Mac) precision~(P) and (wghtd.) recall (R, R$^*$).}
  \label{table-entity-linking-evaluation}
    \begin{footnotesize}    \setlength{\tabcolsep}{4pt}
      \begin{tabular}{@{}l@{$\;$}ccccccc@{}}
       \toprule
                      & Ref.                                  &          MicR  &      MicR$^*$  &          MacR  &       MacR$^*$ &          MicP  &          MacP  \\ 
       \midrule                                                                                                                                              
       \multicolumn{7}{@{}l}{\emph{Entity Linking Tools}}\\                                                                                                  
       \midrule                                                                                                                                              
       Our approach   & --                                    & \textbf{0.838} & \textbf{0.859} & \textbf{0.668} & \textbf{0.670} &         0.035  &         0.126  \\
       Nordlys ER     & \cite{hasibi:2017b}                   &         0.735  &         0.776  &         0.543  &         0.548  &         0.002  &         0.009  \\
       TagMe          & \cite{ferragina:2010}                 &         0.333  &         0.410  &         0.385  &         0.401  &         0.328  &         0.399  \\
       Babelfy        & \cite{moro:2014}                      &         0.320  &         0.398  &         0.383  &         0.398  &         0.293  &         0.289  \\
       Smaph          & \cite{cornolti:2016}                  &         0.314  &         0.390  &         0.399  &         0.413  &         0.431  &         0.463  \\
       Dandelion      & \cite{spaziodati:dandelion}           &         0.302  &         0.373  &         0.414  &         0.428  &         0.431  &         0.500  \\
       Nordlys EL     & \cite{hasibi:2017b}                   &         0.293  &         0.359  &         0.579  &         0.593  &         0.780  & \textbf{0.731} \\
       Dexter         & \cite{ceccarelli:2013}                &         0.267  &         0.332  &         0.359  &         0.372  &         0.481  &         0.462  \\
       FEL            & \cite{blanco:2015}                    &         0.250  &         0.309  &         0.313  &         0.324  &         0.273  &         0.333  \\
       TextRazor      & \cite{textrazor:textrazor}            &         0.216  &         0.265  &         0.372  &         0.380  &         0.511  &         0.445  \\
       Radboud EL     & \cite{vanhulst:2020}                  &         0.213  &         0.263  &         0.498  &         0.507  & \textbf{0.789} &         0.627  \\
       Falcon         & \cite{sakor:2019}                     &         0.204  &         0.251  &         0.226  &         0.234  &         0.397  &         0.368  \\
       Ambiverse      & \cite{hoffart:2011}                   &         0.011  &         0.013  &         0.259  &         0.259  &         0.750  &         0.263  \\
       \midrule                                                                                                                                                    
       \multicolumn{7}{@{}l}{\emph{Entity Recognition Tools}}\\                                                                                                    
       \midrule                                                                                                                                                    
       AWS Comprehend & \cite{amazon:awscomprehend}           &         0.229  &            --  &         0.476  &            --  &         0.604  &         0.616  \\
       MITIE          & \cite{king:mitie}                     &         0.114  &            --  &         0.358  &            --  &         0.797  &         0.463  \\
       Flair NER      & \cite{akbik:2018}                     &         0.129  &            --  &         0.374  &            --  &         0.787  &         0.487  \\
       LingPipe NER   & \cite{aliasi:lingpipe}                &         0.109  &            --  &         0.321  &            --  &         0.497  &         0.410  \\
       DeepPavlov     & \cite{burtsev:deeppavlov}             &         0.048  &            --  &         0.269  &            --  &         0.478  &         0.305  \\
       Stanford NER   & \cite{finkel:2005}                    &         0.011  &            --  &         0.257  &            --  &         0.563  &         0.261  \\
       OpenNLP        & \cite{apache:opennlp}                 &         0.000  &            --  &         0.246  &            --  &         0.000  &         0.246  \\
       \midrule                                                                                                                                                    
       No-Entity Baseline & --                                &         0.000  &         0.000  &         0.246  &         0.246  &         0.000  &         0.246  \\
       \bottomrule
     \end{tabular}
  \end{footnotesize}
\end{table}

\subsection{Query Interpretation Comparison}
Our actually addressed task is not entity linking but entity-based query interpretation on which we also try to compare as many systems as possible. For entity linkers that have no combination step, we apply the GIF~algorithm of \citet{hasibi:2015} that needs linked entities plus mention spans. Hence, entity recognition systems and also Nordlys~ER and Falcon (no mention spans) cannot be tested.

The GIF~algorithm only combines entities but ignores the non-linked terms of a query. For a fair comparison, we post process the GIF~output and add the potentially missing non-linked segments to the interpretation. To not place tools at a disadvantage that also link to concepts besides entities, we manually annotated concepts in the test set and treat them as not-linked segments in the evaluation. A computed interpretation counts as a \emph{complete match} when the segmentation and the contained entities correspond to an interpretation in the ground truth while a \emph{partial match} matches at least the entities of a ground truth interpretation. 

Besides effectiveness, we also measure the efficiency (including~GIF for methods without a combination phase) on a~PC running Ubuntu~20.10 on an AMD\textsuperscript{\textregistered} Ryzen Threadripper 2920x@4.30\,GHz with 128\,GB~RAM (default OS~settings for caches, etc.).

\begin{table}[t]
  \caption{Interpretation results on our test set for interpretations that are at least ``moderately likely''. R, R$^*$, P: (weighted) recall and precision. Time: average time per query.}
  \label{table-interpretation-evaluation}
  \begin{scriptsize}   \setlength{\tabcolsep}{4pt}
    \begin{tabular}{@{}lccccccccr@{}}
      \toprule
      & \multicolumn{4}{c}{Partial Matches} & \multicolumn{4}{c}{Complete Matches} & \multirow{2}{2em}{Time \null\;(ms)} \\
      \cmidrule(l){2-5} \cmidrule(l){6-9}
                   &              R & R* & P & F\textsubscript{1} & R & R* & P & F\textsubscript{1} & \\
      \midrule
      Our approach & \textbf{0.472} & \textbf{0.479} & \textbf{0.506} & \textbf{0.451} & \textbf{0.295} & \textbf{0.301} & \textbf{0.336} & \textbf{0.283} &         47  \\
      Dexter       &         0.306  &         0.311  &         0.392  &         0.337  &         0.230  &         0.235  &         0.312  &         0.246  &        282  \\
      Nordlys EL   &         0.277  &         0.282  &         0.379  &         0.289  &         0.189  &         0.194  &         0.278  &         0.207  &      1,533  \\
      Radboud EL   &         0.223  &         0.224  &         0.289  &         0.236  &         0.144  &         0.145  &         0.199  &         0.155  &        200  \\
      Smaph        &         0.194  &         0.198  &         0.261  &         0.208  &         0.176  &         0.180  &         0.243  &         0.190  &    116,425  \\
      Dandelion    &         0.194  &         0.198  &         0.261  &         0.207  &         0.166  &         0.169  &         0.226  &         0.177  &         74  \\
      TagMe        &         0.176  &         0.181  &         0.228  &         0.187  &         0.165  &         0.169  &         0.216  &         0.175  &         99  \\
      Babelfy      &         0.147  &         0.149  &         0.193  &         0.156  &         0.112  &         0.117  &         0.160  &         0.124  &         49  \\
      TextRazor    &         0.140  &         0.140  &         0.186  &         0.149  &         0.098  &         0.099  &         0.131  &         0.105  &        367  \\
      FEL          &         0.133  &         0.136  &         0.173  &         0.141  &         0.133  &         0.136  &         0.173  &         0.141  & \textbf{22} \\
      Ambiverse    &         0.013  &         0.013  &         0.017  &         0.013  &         0.007  &         0.007  &         0.011  &         0.009  &         53  \\
       \bottomrule
     \end{tabular}
  \end{scriptsize}
\end{table}

Table~\ref{table-interpretation-evaluation} shows the evaluation results on our test set for interpretations that are at least ``moderately likely''. Our approach achieves the by far highest effectiveness and also is faster than all other systems except~FEL. With a partial match recall of~0.47, our approach identifies all entities for almost half of the interpretations while the second best system Dexter (also running completely locally) needs six times more time for its lower 0.31~partial match recall---a recall that our system even almost achieves for the harder complete matches. The third-ranked Nordlys~EL system is slower than most other approaches even though we run it completely locally while the fifth-ranked Smaph~system needs a lot of time for HTTP~requests to several external~APIs.
\section{Conclusions} \label{sec:conclusions}
Our new approach to entity-based query interpretation identifies fitting combinations of important entities at a much higher accuracy and much faster than the previously best systems. With its run time of about~50ms per query (further efficiency tweaks may be applicable), it can already very well serve as a query understanding step in ``production systems'' to determine whether a query has just one interpretation or whether the search results could be diversified. At the same time, the detected entity combinations also help to populate the knowledge box next to the search results.

The core idea of our approach is to combine, for the first time, the query understanding steps of segmentation and entity linking. A fast query segmentation in a pre-processing step helps to substantially reduce the run time of the combination phase of the query interpretation process. Besides the new approach, we also construct our new Webis-QInC-22 dataset of queries annotated with interpretations as well as explicit and implicit entities by combining and extending the publicly available previous query-entity datasets. In a large-scale comparison on the new corpus against the publicly available query--entity linking and interpretation approaches, we show our new approach to achieve a better interpretation accuracy at a better run time than the previously best systems.\footnote{Our corpus, code, and results are publicly available to ensure reproducibility: \url{https://webis.de/data/webis-qinc-22.html}, \url{https://github.com/webis-de/WSDM-22}}

An interesting direction for future work is the inclusion of an actual implicit entity linker---an entity category more or less ignored by the current query interpretation approaches that focus on explicit entities. To this end, we plan to add more queries with implicit entities to our corpus. Additionally, we will also enrich the corpus with related entities for all the queries to form a reusable large-scale dataset for many query--entity-related tasks.

Another interesting direction for future work is to analyze the extent of how any keyword-based ambiguities are transferred or would be resolved in more verbose voice search environments. The verbosity might, for instance, help to avoid ambiguous segmentations and utterance pauses may help to segment voice queries.


%


\end{document}